\documentclass[floatfix,secnumarabic,amssymbpreprint,nobibnotes,nofootinbib,aps,pra,showpacs,superscriptaddress, twocolumn]{revtex4-1}

\usepackage{epsfig}
\usepackage{amssymb}
\usepackage[colorlinks=true, citecolor=red, linkcolor=blue, urlcolor=blue]{hyperref}
\usepackage{times}
\usepackage{amssymb}
\usepackage{amsmath}
\usepackage{cleveref}
\usepackage{mathrsfs}
\usepackage{graphicx}
\usepackage{epsfig}
\usepackage{dcolumn}
\usepackage{color}
\usepackage{xcolor}
\usepackage{bm}
\usepackage{graphics,psfrag}
\usepackage{graphicx,psfrag}
\usepackage{braket}
\usepackage{soul}
\usepackage{subfigure}
\newcommand{\be}{\begin{equation}}
\newcommand{\ee}{\end{equation}}
\newcommand{\bea}{\begin{eqnarray}}
\newcommand{\eea}{\end{eqnarray}}
\newcommand{\ba}[1]{\begin{array}{#1}}
\newcommand{\ea}{\end{array}}

\begin{document}
\title{Zero-threshold correlated-photon laser with a single trapped  atom in a bimodal cavity}
\author{Anushree Dey}
\affiliation{School of Physical Sciences, Indian Association for the Cultivation of Science, Jadavpur, Kolkata 700032, India.}
\author{Arpita Pal}
\thanks{Present address: Department of Optics, Palacký University, 17. listopadu 1192/12, 771 46 Olomouc, Czech Republic.}
\affiliation{Centre for Quantum Engineering, Research and Education,
TCG CREST, Salt Lake, Kolkata 700091, India.}
\author{Subhasish Dutta Gupta}
\email{sdghyderabad@gmail.com}
\affiliation{Department of Physical Sciences, Indian Institute of Science Education and Research (IISER) Kolkata. Mohanpur 741246, India.}
\affiliation{Tata Centre for Interdisciplinary Sciences, TIFRH, Hyderabad 500107, India.}
\affiliation{School of Physics, University of Hyderabad, Hyderabad 500046, India.}
\author{Bimalendu Deb}
\email{msbd@iacs.res.in}
\affiliation{School of Physical Sciences, Indian Association for the Cultivation of Science, Jadavpur, Kolkata 700032, India.}
\begin{abstract}
We demonstrate theoretically the feasibility of correlated entangled photon-pair generation with vanishing threshold in a bimodal cavity setup that uses a single $V$-type three level atom pumped by  dual incoherent sources and driven by two coherent fields. The photon-pair is shown to be entangled only for low levels of the incoherent pumps and owes its origin solely to the coherent drives. Our results show  that the dual incoherent pumping with no coherent drive can lead to amplification of the cavity fields with strong inter-mode antibunching but no entanglement. Though only coherent drives with no incoherent pumping can produce entangled photon-pairs, the entangled cavity fields can not be amplified   beyond a certain limit using only coherent drives. However, the use of even small incoherent pumping in the presence of the coherent drives   can amplify the generated entangled photon-pairs significantly. We analyse our results in terms of an interplay between coherent and incoherent processes involving cavity-dressed states. Both the inter- and intra-mode HBT functions exhibit temporal oscillations in the strong-coupling cavity QED regime.  Our theoretical scheme for the generation of nonclassical and entangled photon pairs may find interesting applications in quantum metrology and quantum information science.
\end{abstract}
\maketitle

\section{Introduction}
Over the years, cavity quantum electrodynamics (CQED) \citep{berman:cqed, walther2006cavity,miller2005trapped} where atoms interact with quantized electromagnetic fields have emerged as an excellent platform  for engineering entanglement \citep{RevModPhys.73.565, PhysRevLett.90.027903}, photon blockade \citep{PhysRevLett.79.1467,Nature2,PhysRevA.95.063842}, generation of quantum light \cite{rempe:prl124:093603:2020,Mukamel:jpb:2020} and a variety of quantum correlations \citep{PhysRevLett.80.3948,PhysRevLett.107.023601}. Antibunching and entanglement are a crucial resource for quantum information processing, metrology and quantum teleportation \citep{PhysRevA.98.012121,PhysRevA.99.023828}. In particular, correlated or entangled photons are essential for photon-based quantum metrology and quantum technology \citep{Nature4,PhysRevLett.118.133602,PhysRevA.100.053802,PhysRevLett.94.023601,PhysRevLett.127.123602,PhysRevA.97.023822}. One of the important sources of correlated photons used  for a variety of fundamental studies  over the years is optical parametric oscillator (OPO) \citep{mandel:prl59:2044:1987,kwait:prl75:4337:1995,RevModPhys.84.777.2012,muller:NaturePhotonics8:224:2014,PhysRevA.103.053710,PhysRevA.62.033802}. However, the generation of correlated photons by an OPO is probabilistic. In recent times, several theoretical proposals for generating nonclassical or entangled radiation fields by coherently driving multilevel atoms inside a cavity have been put forward \cite{xion:prl94:023601:2005, zubairy:pra77:062308:2008,fang:pra81:012323:2010}. Lasing into a single-mode cavity field due to coherent as well as incoherent pumping of multi-level atoms in strong-coupling CQED regime has also been studied \cite{Devi:pra}. Although, with the currently available conventional laser technology, it is possible to produce extremely narrow-band and high-quality single-mode coherent light sources  there hardly exists any standard technique to create a laser source that can deliver  entangled or non-classically correlated photons on demand. 

A one-atom laser using CQED  was theoretically conceived three decades ago \cite{muandsavage:pra:1992}. Since then several theoretical as well as experimental studies \cite{ritsch:jmo41:609:1994,walther_PhysRevA.55.3923,kimble_mckeever2004deterministic,khajavikhan2012thresholdless,gsa-sdg:pra42:1737:1990,walther:epl:1997,Nature1,PhysRevA.70.023814} on single-atom CQED laser have been carried out. The steady-state properties of an incoherently pumped single-atom in a single-mode cavity reveal vanishing threshold for lasing action of the cavity mode with interesting atom-photon  and photon-photon correlations \cite{gsa-sdg:pra42:1737:1990}. A  laser with a single trapped ion inside a high-Q cavity was theoretically proposed by Meyer, Briegel and Walther \cite{walther:epl:1997} about two decades ago. 1980s and 1990s had witnessed a lot of  progress  in the development of micromaser \cite{PhysRevA.34.3077_filipowicz,Haroche_micromaser_PhysRevA.36.3771,rempe_micromaser_PhysRevLett.64.2783,walther_micromaser_PhysRevLett.82.3795} that uses a mono-energetic  beam of Rydberg atoms passing through a high-Q cavity one by one. An optical analog of micromaser, namely microlaser had been realized in the 90s \cite{martini_microlaser_PhysRevA.46.4220,an_microlaser_PhysRevLett.73.3375}. Although  a micromaser or microlaser makes use of the strong atom-cavity interactions of single atoms, it is not  truly a one-atom maser or laser as it  results from  the cumulative effect of interactions of a beam of atoms with the cavity.
The recent advancement of trapping and cooling of single atoms and ions opens new  prospects for developing a microscopic laser with a single trapped atom inside a high-Q cavity for controlled generation of nonclassical light.  
In this context, McKeever  {\it et. al.}  \citep{Nature1,PhysRevA.70.023814} had experimentally demonstrated  lasing with a single trapped atom  in the strong-coupling CQED regime  in 2003. A one-atom  CQED laser is fundamentally different from a standard laser in many respects. First, it can operate at vanishingly small or zero threshold. Second, its frequency depends on the atom-cavity coupling. Third, it can be operated with a single trapped atom at a few photon levels enabling substantial microscopic control over the output signal amplitude and photon statistics. Fourth, as shown in Ref. \cite{Nature1}, it can produce antibunched light which is nonclassical. 
 
\begin{figure}
\centering
\includegraphics[width=1\linewidth]{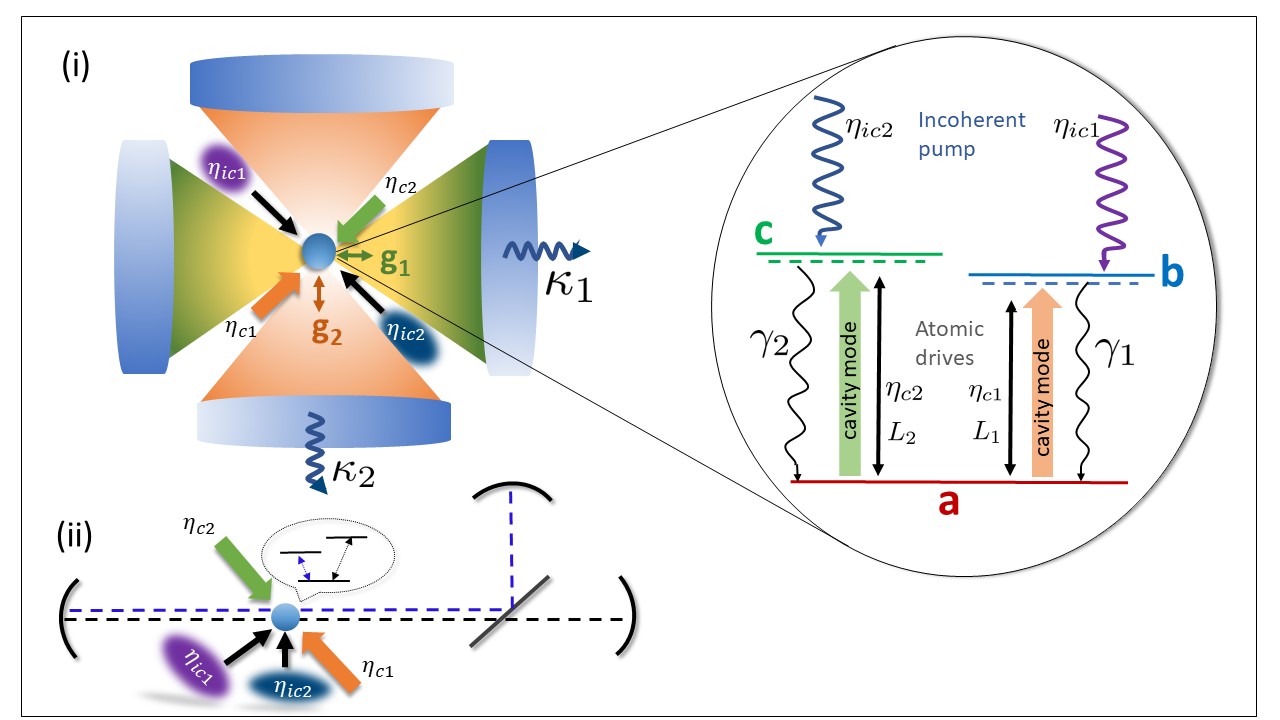}
\caption{A schematic diagram showing a single trapped $V$-type three-level atom interacting with two cavity modes either in a crossed cavity (i) or a doubly-resonant cavity (ii) setup. The zoomed view shows the atomic level diagram with $a$, $b$ and $c$ denoting the three atomic levels and the arrows indicate various transition pathways. The cavity modes 1 and 2 couple the lowest level $a$ to the two upper levels $b$ and $c$, respectively. Here $\eta_{ic1}$ and $\eta_{ic2}$ represent two incoherent pumping rates  while $\eta_{c1}$ and $\eta_{c2}$ are the coherent coupling rates due to  laser fields $L_1$ and $L_2$ that drive the corresponding atomic transitions. The incoherent pumping may be accomplished by exciting the atom  from the level $a$ to a higher level (not shown in the diagram) followed by spontaneous decay to either level $b$ or $c$. $\gamma_1$ and $\gamma_2$ are the atomic decay rates for the transitions $b \rightarrow a $ and $c \rightarrow a$, respectively;  while $\kappa_1$ and $\kappa_2$ are the cavity decay rates for the modes 1 and 2, respectively. }
\label{fig 1}
\end{figure}

Here we propose a model for a single-atom threshold-less bimodal correlated-photon laser (CPL) based on strong-coupling CQED. Our proposed laser system is schematically shown in Fig.\ref{fig 1}. It may  consist of  a pair of crossed cavities or a bimodal cavity, a single trapped $V$-type three-level atom inside  the bimodal cavity or at the symmetry point of the crossed cavities, two incoherent pumps and two lasers to drive the two atomic transitions. The two excited levels of the atom are incoherently  pumped by external sources while a pair of coherent drives (lasers) couple the atomic ground state with the two excited states.

We first discuss the effects of dual incoherent pumping without any coherent drive and then we focus on the effects of dual coherent fields that drive two atomic transitions. 
It is well-known that driving a three-level atom with only coherent drives can lead to coherence between the atomic levels \cite{PhysRevA.82.063837, PhysRevA.84.063801,rakshit2014decay} which is central to  many quantum interference effects \cite{ficek2005quantum} and nonclassical correlations \cite{PhysRevA.82.063837,PhysRevA.84.063801}. However, to the best of our knowledge, it is not known how dual incoherent pumping can influence quantum dynamics of a three-level emitter in a bimodal cavity. Only dual incoherent pumping of a $V$-type three-level system in free space can not produce any excited-state atomic coherence unless the two transition dipole moments are non-orthogonal \cite{scully_PhysRevA.74.063829}. In our model,  the two transition dipole moments are orthogonal, since  all three levels of the atom  have different angular momenta.   Nevertheless, the fact that our results show nonclassical field-field correlations even with only incoherent pumps has to do with an interplay between strong coupling CQED and the incoherent pumping as we elaborate later.

Here we demonstrate an interesting interplay between coherent and incoherent processes within the framework of strong-coupling cavity-QED. The  incoherent pumping leads to the amplification of the two cavity modes while bimodal cavity field-induced coherent processes result in nonclassical correlations.  With dual incoherent pumping only, the system produces inter-mode photon antibunching  with amplifications in both the cavity fields. This nonclassical correlation can be attributed to an interplay between the coherent Rabi dynamics due to the cavity fields and the dual incoherent pumping. We show that the relative intensity of the two cavity fields, intra-mode and inter-mode photon statistics can be controlled by tuning the relative strength of the dual incoherent pumps. Particularly interesting is the situation where either of the two excited states is metastable and the other is short-lived. In this situation, depending on the relative pump strength, the cavity mode that couples to the metastable state exhibits nonclassical photon statistics while the other mode coupled to the short-lived state shows classical photon statistics. The intra-mode photon-photon correlation is always nonclassical. Thus dual incoherent pumps can act as a knob or switch to control bimodal photon-photon correlations. 

 Apart from the effects of dual incoherent pumping, we  study the effects of dual coherent driving of the atom on the properties of the cavity fields. Our results show that when the atom is driven by external coherent drives in the absence of incoherent pumping or in the presence of small incoherent pumping,  the two cavity modes become entangled. We show that the origin of the photon-photon entanglement lies in the dressing of the cavity-dressed states by the external coherent fields. When the photons are emitted from such doubly dressed states, they carry forward the nonclassical properties of the dressed states in the form of inter-mode photon-photon entanglement.  Only coherent drives can not amplify the fields beyond the saturation level at which 50\% population lies at the atomic ground state and the rest 50\% is distributed  between the two excited states  \cite{scully_PhysRevA.74.063829}. With only a single emitter, even at saturation level with coherent drives only, the output cavity signal will be feeble. So, the role of small incoherent pumping in addition to the coherent drives is important as it can lead to amplification of the entangled photons.

The paper is organised in the following way: we present theoretical formulation of our problem in Sec.~\ref{sec2}. We discuss and analyse our results in Sec.~\ref{sec3}. The paper is concluded in Sec.~\ref{sec4}.

\section{Theoretical formulation}\label{sec2}
We consider a trapped $V$-type three-level atom either at the intersection of a pair of crossed cavities as schematically shown Fig.~\ref{fig 1} (i) or in a doubly-resonant cavity setup as shown in  Fig.~\ref{fig 1} (ii) . Here $a$ denotes the ground state, and $b$ and $c$ represent the two excited states which may be degenerate or quasi-degenerate. The two cavity modes 1 and 2 are tuned to the transitions $a \rightarrow b$ and $a \rightarrow c$, respectively. The two excited levels $b$ and $c$ are pumped  by dual incoherent sources while the two atomic transitions may be driven by two classical laser fields as depicted in Fig.~\ref{fig 1}.

\subsection{Hamiltonian}
The Hamiltonian for this system is 
\begin{equation}
{\cal H} =  {\cal H}_0 + {\cal H}_{int}+ {\cal H}_{drive}~,
\end{equation} 
where ${\cal H}_0 = \hbar \omega_b \mid b\rangle \langle b \mid+ \hbar \omega_c \mid c\rangle \langle c \mid + \hbar \omega_1 a^{\dag}_1 a_1 + \hbar \omega_2 a^{\dag}_2 a_2~$ is the free Hamiltonian with $\omega_b$ and $\omega_c$ being the respective atomic transition frequencies. We choose the energy of the lowest level $a$ to be zero. Here $a_i (a_i^{\dagger})$ denotes the annihilation (creation) operator of the $i$-th (=1,2) cavity mode. Under electric-dipole and rotating-wave approximation (RWA), the interaction Hamiltonian ${\cal {H}}_{int}$ can be written as 
\begin{eqnarray}
 {\cal {H}}_{int} = \hbar \left( g_1 a_1 \sigma^{\dagger}_1 + g_2 a_2 \sigma^{\dagger}_2\right) + {\rm H.c.}~,
\end{eqnarray}
where $g_1$ and $g_2$ are the atom-cavity coupling constants that depend on dipole moment and the respective transition field amplitude. $\sigma^{{\dagger}}_{1}= \mid b \rangle \langle a \mid $ and $\sigma^{{\dagger}}_{2} =  \mid c \rangle \langle a \mid$ represent the atomic raising operators and the corresponding lowering operators are  $\sigma_{1}= \mid a \rangle \langle b \mid $ and $\sigma_{2} = \mid a \rangle \langle c \mid$, respectively. Here the Hamiltonian corresponding to the driving of  the atomic transitions by two classical fields  is given by  
\begin{equation} 
 {\cal H}_{drive} = \hbar \eta_{c1} \sigma_1 e^{i \omega_{L_1}t}  + \hbar \eta_{c2} \sigma_2 e^{i \omega_{L_2}t}  + {\rm H.c.}~, 
\end{equation} 
where $\eta_{c1}$ and $\eta_{c2}$ are the laser-atom coupling constants. In the rotating reference frame of the laser frequencies, the Hamiltonian reads as
\bea
{\cal H} &=& \hbar \left  (\Delta_b \mid b\rangle \langle b \mid + \Delta_c \mid c\rangle \langle c \mid +  \delta_{1L} a^{\dag}_1 a_1 +  \delta_{2L} a^{\dag}_2 a_2 \right ) \nonumber\\
&&+{\cal H}_{int} +  \hbar \left [ \eta_{c1} \left (\sigma_1 + \sigma^{\dag}_1 \right ) +  \eta_{c2} \left  (\sigma_2 + \sigma^{\dag}_2 \right ) \right ]~,
 \eea
 where the detuning parameters are $\Delta_b = \omega_b - \omega_{L_1}$, $\Delta_c = \omega_c - \omega_{L_2}$, $\delta_{1L} = \omega_1 - \omega_{L_1}$ and $\delta_{2L} = \omega_2 - \omega_{L_2}$.  Here $\omega_{L_1}$ and $\omega_{L_2}$ are the frequencies of $L_1$ and $L_2$ lasers. The detuning of the two cavity fields are denoted by $\delta_1 = \omega_1 - \omega_b$ and $\delta_2 = \omega_2 - \omega_c$.

\subsection{Dressed-states} 
 In the absence of all incoherent processes, one can develop a dressed state description of the driven three-level atom interacting with two cavity modes. Dressed state picture of a CQED problem helps to gain insight into the physical processes underlying the relevant quantum dynamics, particularly in the strong-coupling regime.  
 We use the two-mode Fock basis $\mid n_1, n_2 \rangle$ for the two cavity fields, where $n_1=0, 1, 2, \cdots, \infty $ and $n_2 = 0, 1, 2, \cdots, \infty$ denote the number of photons in the mode 1 and 2, respectively. One can construct three bare basis states $\mid a, n_1, n_2 \rangle$, ~ $\mid b, n_1-1, n_2 \rangle$,~ ${\rm and} \mid c, n_1, n_2-1 \rangle$ which will be involved in the closed coherent quantum dynamics in the absence of all incoherent  processes and couplings to the classical derives.

Defining the excitation number operators $N_1 = a^{\dag}_1 a_1 + \mid b \rangle \langle b \mid$ and $N_2 = a^{\dag}_2 a_2 + \mid c \rangle \langle c \mid$ and applying the  unitary transformation 
$\hat{U}(t) = {\rm exp} \left[-i \left(\omega_1 N_1 + \omega_2 N_2\right)t\right]~$\cite{barnett2002methods}, one can  obtain the following  effective  Hamiltonian in the absence of the lasers  
\bea
\tilde{H} = -\delta_1 \mid b \rangle \langle b \mid - \delta_2 \mid c \rangle \langle c \mid + {\cal H}_{int}~.
\label{Heff}
\eea
If the two cavity fields are tuned to the two-photon resonance  $\delta_{1} = \delta_{2} = \delta$, then the eigenvalues of $\tilde{H}$ are $\tilde{E}_0 = -  \delta $ and $\tilde{E}_{\pm} = - \frac{\delta}{2} \pm  \frac{1}{2} \sqrt{ \delta^2 + 4 \left ( n_1 g_1^2 + n_2 g_2^2 \right )}$. For limited Fock states, the characteristic eigenvalue and eigenstate analysis can be found elsewhere \citep{Pal2019PhotonphotonCW}.

The eigenstate $\mid \psi_0(n_1,n_2) \rangle$ corresponding to the eigenvalue  $\tilde{E}_0$  is given by 
\bea
\begin{split}
\mid \psi_0(n_1, n_2) \rangle 
= \frac{1}{\sqrt{g_2^2n_2+g_1^2n_1}} [ g_1\sqrt{n_1}\mid c,n_1,n_2-1\rangle &\\
- g_2\sqrt{n_2}\mid b,n_1-1,n_2\rangle ] ~.&
\end{split}
\eea
which is a coherent superposition of the two atom-field joint states involving only the two atomic excited states. It results from the destructive quantum interference between the two cavity field-induced transition pathways $ \mid b,n_1-1,n_2\rangle \longleftrightarrow \mid a, n_1, n_2 \rangle $ and  $ \mid c,n_1,n_2 - 1 \rangle \longleftrightarrow \mid a, n_1, n_2 \rangle $ connecting 
the two joint atom-field excited states to the joint atom-field state  with the atom being in the ground state. This is an excited dark state due to the quantized cavity fields.  The dark-state resonance and associated phenomena such as coherent population trapping and electromagnetically induced transparency (EIT)  are well-known for a Lambda-type three level system interacting with two classical coherent fields. In a recent work, Rempe's group \cite{rempe:prl124:093603:2020}   has experimentally demonstrated 
the generation of single-mode quantum light utilizing the dark state of a coherently driven $\Lambda$-type three-level atom inside a single-mode cavity.  
However, for a $V$-type three-level system, the dark state resonance and associated effects are qualitatively different \citep{Nature3}.  The excited-state coherence in $V$-type three-level atoms driven by classical coherent fields may lead to transient coherent population trapping \citep{ARIMONDO1996257}, quantum beat  and correlated two-mode lasers as shown by Scully's group \cite{scully:prl55:2802:1985,scully:prl60:1988,scullu:zubairy:pra35:752:1987,scully:pra38:1988}.

The other two dressed states corresponding to the eigenvalues $E_{\pm}$ are coherent superposition of all three bare states. For $\delta = 0$, these two dressed states can be expressed as 
\bea
\begin{split}
\mid \psi_{\pm}(n_1, n_2) \rangle = \frac{1}{\sqrt{2}} \mid a,n_1,n_2\rangle \pm  \frac{1}{\sqrt{2(g_2^2n_2+g_1^2n_1)}} &\\
\times \left [ g_2\sqrt{n_2}\mid c,n_1,n_2-1\rangle + g_1\sqrt n_1\mid b,n_1-1,n_2\rangle \right ]~.&
  \end{split}
  \label{eqn psipm}
\eea
We call these three dressed states, namely, $\mid \psi_{0}(n_1, n_2)$, $\mid \psi_{\pm}(n_1, n_2) \rangle $ as belonging to the $(n_1, n_2)$-photon sector.  Thus the two-mode strong-coupling CQED with a $V$-type three-level atom enables the creation of tripartite superposition states involving the atom and the two quantized fields. The incoherent pumping and other incoherent processes such as spontaneous emission as well as classical drives will induce transitions to the states outside these three dressed  states, thus coupling all other photon sectors. While dual incoherent pumping can lead to the enhancement of the excited-state population and the consequent lasing into the cavity modes, the coherent atomic drives can generate further coherence between the cavity-dressed states. This coherence  plays an essential role in creating entanglement between two cavity modes.

\begin{figure*}
\includegraphics[width=0.9\linewidth]{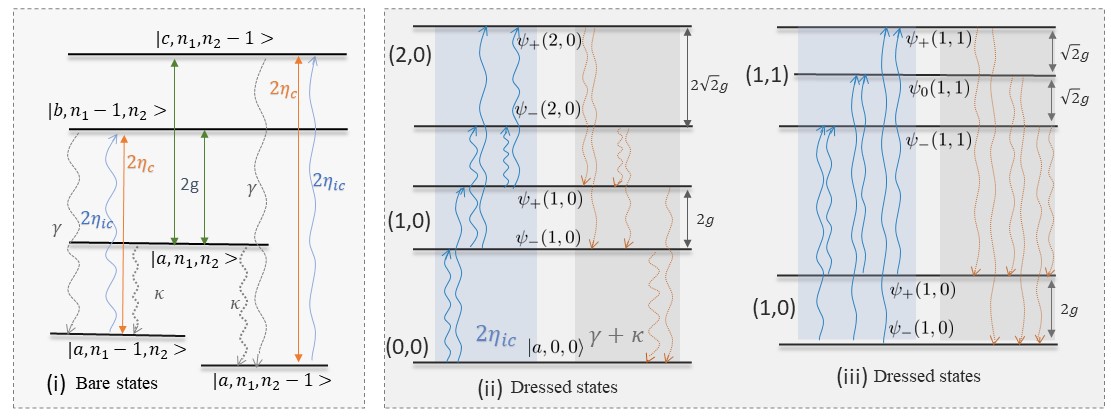}
\caption{{ (i) Bare-state level diagram with various transition pathways due to different incoherent processes and 
coherent atomic drives. (ii) and (iii) A few low lying dressed states with transitions due the incoherent processes only. 
Here $g_1= g_2 = g$ (atom-field coupling), $\eta_{ic1} = \eta_{ic2} = \eta_{ic}$ (incoherent pumps) and $\eta_{c1} = \eta_{c2} = \eta_c$ (coherent drives). In (i), the green and orange- double headed arrows represent the coherent couplings due to the cavity and driving fields, respectively. The curvy upward and downward arrows indicate incoherent pumping and atomic or cavity decay processes, respectively. In the dressed-state picture, the frequency spacing between the dressed states belonging to the same photon sector are marked. Note that the dressed-state diagram involving couplings among (0,0) $\Longleftrightarrow$ (0,1) $\Longleftrightarrow$ (0,2)  is the same as in (ii) with `(1,0)' and `(2,0)' being replaced by `(0,1)' and `(0,2)', respectively. Similarly, the dressed-state diagram involving (0,1)$\Longleftrightarrow$ (1,1), photon sectors will be same as in (iii) with  `(1,0)' being replaced by `(0,1)'. The spacing between different levels is not to be scaled.} }
\label{fig 2}
\end{figure*}

To keep the dressed-state analysis simple, we assume that $\delta = 0 $ and $g_1 = g_2 = g$, $\eta_{ic1} = \eta_{ic2} = \eta_{ic} $, $\kappa_1 = \kappa_2 = \kappa$ and $\gamma_1 = \gamma_2 = \gamma$.  The (0,0)-photon sector, that is when both the cavity fields are in vacuum, there exists only one joint state $\mid a, 0, 0 \rangle$. Each of the $(n_1,0)$- and $(0,n_2)$-photon sectors have two dressed states
\begin{eqnarray}
 \mid \psi_{\pm}(n_1,0) \rangle &=& \frac{1}{\sqrt{2}} \left [ \mid a, n_1, 0 \rangle \pm \mid b, n_1-1,0 \rangle \right ]~,
\nonumber \\
 \mid \psi_{\pm}(0,n_2) \rangle &=& \frac{1}{\sqrt{2}} \left [ \mid a, 0, n_2 \rangle \pm \mid c,0,n_2-1 \rangle \right ].
 \label{eqn psibc}
\end{eqnarray}
The corresponding eigenvalues are $\pm g$. These are the familiar dressed states of two-state Jaynes-Cummings model. If the fields contain zero photon in one mode and a finite number of photons in the other mode, then in the absence of incoherent processes and any classical drive, the system effectively reduces to a two-level system. However, atomic drives and incoherent processes can connect $\psi_{\pm}(n_1,0)$ and $\psi_{\pm}(0,n_2)$ to  $\mid a, n_1 - 1,0 \rangle$ and $\mid a, 0,n_2 - 1\rangle$, respectively.  As we will show in the next section, when dual incoherent pumps are weak, that is, $\eta_{ic} < (\kappa + \gamma) $, the lasing actions and the nature of two-photon correlations between the generated photons can be interpreted in terms of an interplay between the coherent and incoherent transitions involving some low-lying bare or dressed states. For strong incoherent pumping $\eta_{ic} \ge  (\kappa + \gamma) $, higher photon sectors get involved in the dynamics.  The possible transitions between the bare states as well as between the dressed states belonging to a few low-lying photon sectors are schematically shown in Fig.\ref{fig 2}  

\subsection{ Incoherent pumping and damping}
To include incoherent pumping and various damping processes in the system, we resort to the density matrix equation in Lindblad form
\bea
\begin{split}
 \frac{\partial{\hat\rho_{sf}}}{\partial t} = - \frac{i}{\hbar}\left[ {\cal{H}}(\tilde{H}),\hat\rho_{sf}\right] + \sum_{i=1,2}\frac{\kappa_i}{2}\mathcal{L}_{a_i}  + \sum_{j=1,2}\frac{\gamma_{j}}{2}\mathcal{L}_{\sigma_j} &\\
 + \sum_{j=1,2}\frac{\eta_{icj}}{2}\mathcal{L}_{\sigma_j^{\dagger}} ~,&
 \end{split}
 \label{eq1}
\eea
Where $\rho_{sf}$ is the density matrix of joint system, comprising of the sub-system atom (denoted by `$s$' and the cavity fields (denoted by `$f$'). $\cal L_{\rm{x}}$ is  a Liouville super operator where $x$ stands for any of the three operators $a_i,\sigma^{\dagger}_j,\sigma_j$. $\kappa_i$ denotes the decay rate of the cavity field mode $i$ (i = 1, 2). Here the Lindbladian superoperators 
\begin{align*}
\mathcal{L}_{a_i}&=\left[2a_i\rho_{sf}a_i^\dagger- \left \{ a_i^\dagger a_i, \rho_{sf} \right \}\right]~, \\
\mathcal{L}_{\sigma_j}&= \left[ 2\sigma_j\rho_{sf}\sigma_j^{\dagger} - \left \{ \sigma_j^{\dagger} \sigma_j, \rho_{sf} \right \} \right ]~, \nonumber 
 \end{align*}
 describe the cavity and atomic damping, respectively;  while 
 \begin{equation}
\mathcal{L}_{\sigma_j^{\dagger}}= \left[ 2\sigma_j^{\dagger}\rho_{sf}\sigma_j - \left \{ \sigma_j \sigma_j^{\dagger}, \rho_{sf} \right \} \right ]~,  
\end{equation}
describes incoherent pumping to the two upper levels of the atom. Here $\{ \alpha, \beta \} $ implies anticommutation  between $\alpha$ and $\beta$. 
To numerically solve \citep{johansson:qutip:2012, JOHANSSON20131234} Eq.\ref{eq1}  we use joint atom-field basis states $\mid \alpha, n_1, n_2\rangle$ where $\alpha$ denotes any of the three atomic states, $n_1$ and $n_2$ are the photon numbers. For a given set of system parameters, we first verify the convergence of our numerical results against the variation of the photon numbers in the two modes. Once we ensure the convergence, we truncate the Fock basis of the two fields for the chosen parameter set. We numerically  obtain the steady-state solution of the joint atom-field density matrix.

\begin{figure}[h]
\includegraphics[width=0.7\linewidth]{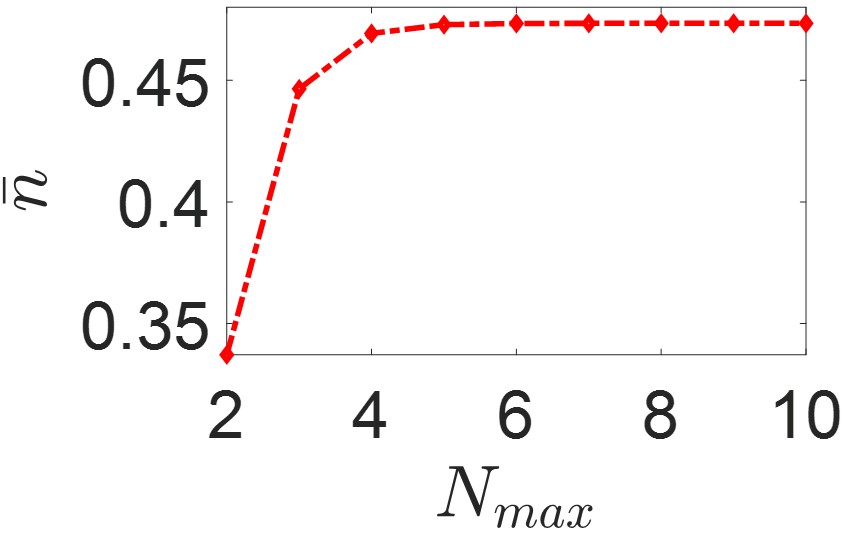}
\caption{The steady-state average photon numbers $\bar n_1=\bar n_2=\bar n$ as a function of the maximum Fock-state number  $N_{max}$ in each mode. The atomic damping rates are $\gamma_1 = \gamma_2 = \gamma$,  the atom-cavity coupling parameters are ${g}_1= {g}_2 = 10 \gamma $, the cavity damping rates are  $\kappa_1=\kappa_2 = \gamma$, the incoherent pump strengths are $ \eta_{ic1}=\eta_{ic2} = 2 \gamma$ and $\delta_1 = \delta_2 = 0$. }
\label{fig 3}
\end{figure}

\subsection{Second-order photon-photon correlations and entanglement}\label{D}
We calculate the expectation value of an observable operator $\hat{O}$ of interest in the  steady-state by
\begin{eqnarray}
 \langle\hat O\rangle\nonumber={\rm Tr}\left[\hat O \hat\rho_{sf}^{ss}\right]&=&\sum_{\alpha}\sum_{n}\sum_{m}\langle \alpha,n,m|\hat O \hat\rho_{sf}|\alpha, n,m\rangle~.
 \label{eq12} 
\end{eqnarray}
where $\alpha$ stands for an atomic state, $n$ and $m$ represent photon number states for mode 1 and 2, respectively. Here $\hat\rho_{sf}^{ss} = \hat\rho_{sf}(t \rightarrow \infty)$ stands for the steady-state joint density matrix. We calculate the two-time Hanbury Brown-Twiss (HBT) second-order correlation function \cite{JURCZAK1995480} as
\bea
g_{ij}^{(2)}(\tau) &=& \frac{\langle a^{\dag}_i(t) a^{\dag}_j(t+\tau)a_j(t+\tau)a_i(t)\rangle}{\langle  a^{\dag}_i(t) a_i (t)\rangle \langle a^{\dag}_j(t+\tau) a_j (t+\tau)\rangle}~.
\label{twotime}
\eea
where $i=1,2$ and $j=1,2$ denote the mode indexes. If $i=j$ then we have intra-mode correlation, otherwise we obtain inter-mode correlation. Here  $\langle a^{\dag}_i a_i\rangle$ is the mean  photon number of the $i^{th}$ cavity mode. Using quantum regression theorem \cite{PhysRev.129.2342,PhysRev.157.213,carmichael1999statistical}, HBT function  can be calculated for  fluctuations around the steady state as shown in Ref.\citep{Pal2019PhotonphotonCW}. 
For $\tau \sim 0$, the system's response manifests into photonic correlation ($g_{ij}^{(2)}(0) > 1$) or anticorrelation ($g_{ij}^{(2)}(0) < 1$). When $g_{ij}^{(2)}(0) \gg 1$, the photons are highly correlated in bunched form. i.e. bunching occurs. In contrast, when $g_{ij}^{(2)}(0) < 1$, there occurs photon antibunching and as $g_{ij}^{(2)}(0)\rightarrow 0$, the perfect photon blockade \citep{PhysRevLett.79.1467} takes place. Now, we define the frequency domain of correlation by Fourier transforming $g_{ij}^{(2)}(\tau)$
\begin{equation}
\mathcal{F}_{ij}(\omega)=\int_{-\infty}^{\infty} [{g_{ij}^{(2)}(\tau)-g_{ij}^{(2)}(\infty)}]e^{(i\omega\tau)} \,d\tau ~.
\end{equation}
Depending on the coherence in the system, $g_{ij}^{(2)}(\tau)$ as a function of $\tau$ may exhibit oscillatory behavior. In the limit $\tau \rightarrow \infty$, $g_{ij}^{(2)}(\tau) \rightarrow 1$. To know the frequencies of the oscillations and other frequency-domain features of HBT correlations \cite{kingshuk_PhysRevA.103.033310}, it is important to study not only the time dependence of $g_{ij}^{(2)}(\tau)$ but also the frequency dependence of $\mathcal{F}_{ij}(\omega)$. In the present context, the HBT correlations as a function of $\tau$ will decay to the steady value of unity  with a time constant $\tau_d \sim (\kappa + \gamma)^{-1}$ while the oscillations will bear the signature of quantized Rabi dynamics.  

A bipartite system is called entangled if its wave function is not product separable into the wave functions of its two subsystems. 
The inseparability criterion of Peres \cite{PhysRevLett.77.1413} and Horodecki \cite{horodecki1996separability} states that the necessary and sufficient condition for the inseparability of the wave function of a bipartite system is the negativity of at least one of the eigenvalues of the partial transpose  of the density matrix of the system. There exist different criteria \cite{PhysRevLett.84.2722,PhysRevLett.84.2726} for testing the existence of bipartite entanglement in terms of Gaussian variables. In the present context, the two cavity modes and the atom may be in a tripartite entangled state. Then to discuss the entanglement between the two modes only, we first reduce the full atom-field density matrix to a reduced matrix of the two modes of cavity fields only by tracing over the atomic states. This reduced field density matrix is the density matrix of our bipartite system. Since the cavity fields are in general in non-Gaussian states, we do not use the criteria of the Refs.\cite{PhysRevLett.84.2722,PhysRevLett.84.2726}. Instead  we make use of the
Peres-Horodecki criterion to test the existence of entanglement between the two cavity modes. 
We take partial transpose  over the mode 2 or 1 of the two-mode reduced field density matrix.   The resulting matrix is denoted as $\rho^{{\rm PT,2}}_f$ or $\rho^{{\rm PT,1}}_f$, respectively. Partial transpose over field mode 2 implies that the basis of the bipartite density matrix change as $\mid mn\rangle \langle \mu \nu \mid \rightarrow \mid m \nu \rangle \langle \mu n\mid$. The negativity of at least one eigenvalue of $\rho^{{\rm PT,2}}_f$ or $\rho^{{\rm PT,1}}_f$ implies that the two field modes are entangled. 

\section{Results and Discussions}\label{sec3}
To present and discuss the results in a systematic way, we first present the results in the absence of classical drives, and analyze the
effects of the dual incoherent pumps. We then discuss the effects of coherent drives in the presence as well as in absence of incoherent pumps.
Thus we clearly discern the effects of coherent drives from the incoherent pumps and vice versa.  In all our numerical results, all frequency quantities are given in unit of $\gamma_2$.

\subsection{The effects of incoherent pump only}
\begin{figure}
\centering
\vspace{.25in}
  \begin{tabular}{@{}cc@{}}
\includegraphics[width=1\linewidth]{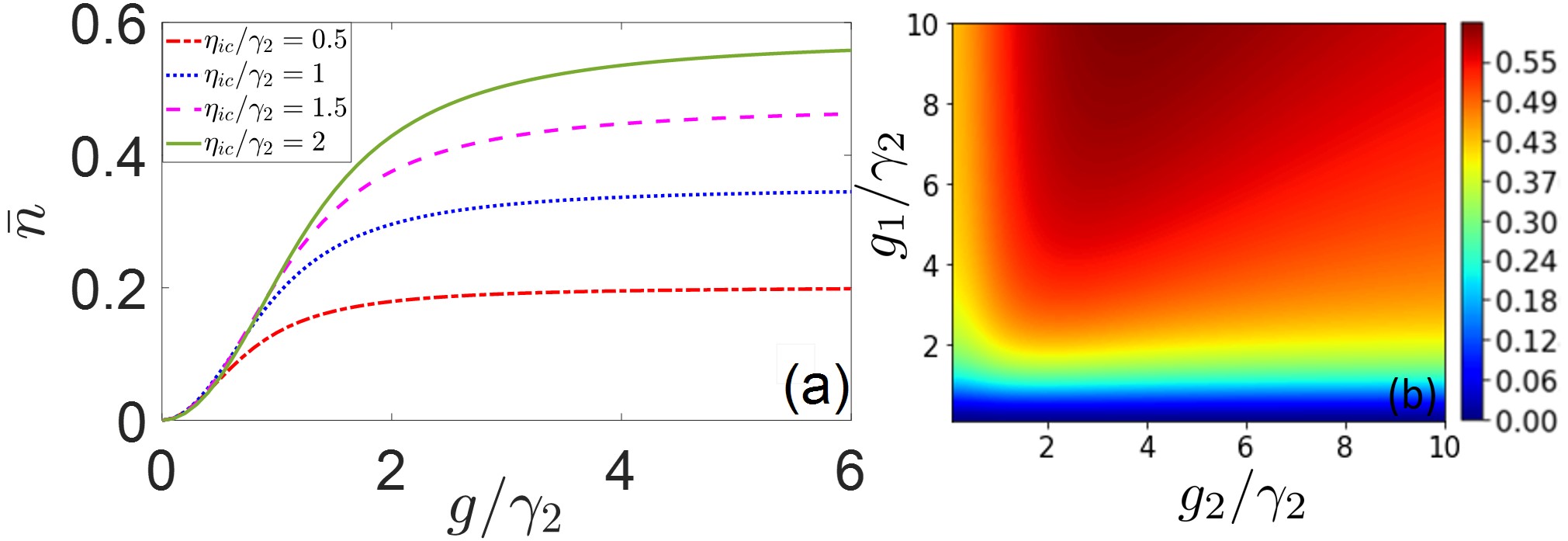}\hfill
     \end{tabular}
\caption{(a) The average number of photons $\bar n_1=\bar n_2=\bar n$ is plotted as a function of scaled atom-cavity coupling parameter ${g}_1/\gamma_2={ g}_2/\gamma_2 = g/\gamma_2$. The other parameters are $\delta_1 = \delta_2 =0$, $\gamma_1 = \gamma_2$, $\kappa_1 = \kappa_2 = \gamma_2$,  $ \eta_{ic1}=\eta_{ic2} = \eta_{ic} =  2 \gamma_2$ (solid), $\eta_{ic} =  1.5 \gamma_2$ (dashed), $\eta_{ic} = \gamma_2$ (dotted),  $\eta_{ic} = 0.5 \gamma_2$ (dashed-dotted). The surface plot (b) depicts the variation of $\bar n_1$  as a function of  ${ g}_1/\gamma_2$ and ${g}_2/\gamma_2$, for $ \eta_{ic1}=\eta_{ic2} = 2 \gamma_2$ with all other parameters remaining fixed. The surface plot of $\bar n_2$ will be  same as in (b) but with $g_1$- and $g_2$-axis being exchanged. }
\label{fig 4}
\end{figure}

\begin{figure}
\centering
\vspace{.25in}
  \begin{tabular}{@{}cc@{}}
\includegraphics[width=1\linewidth]{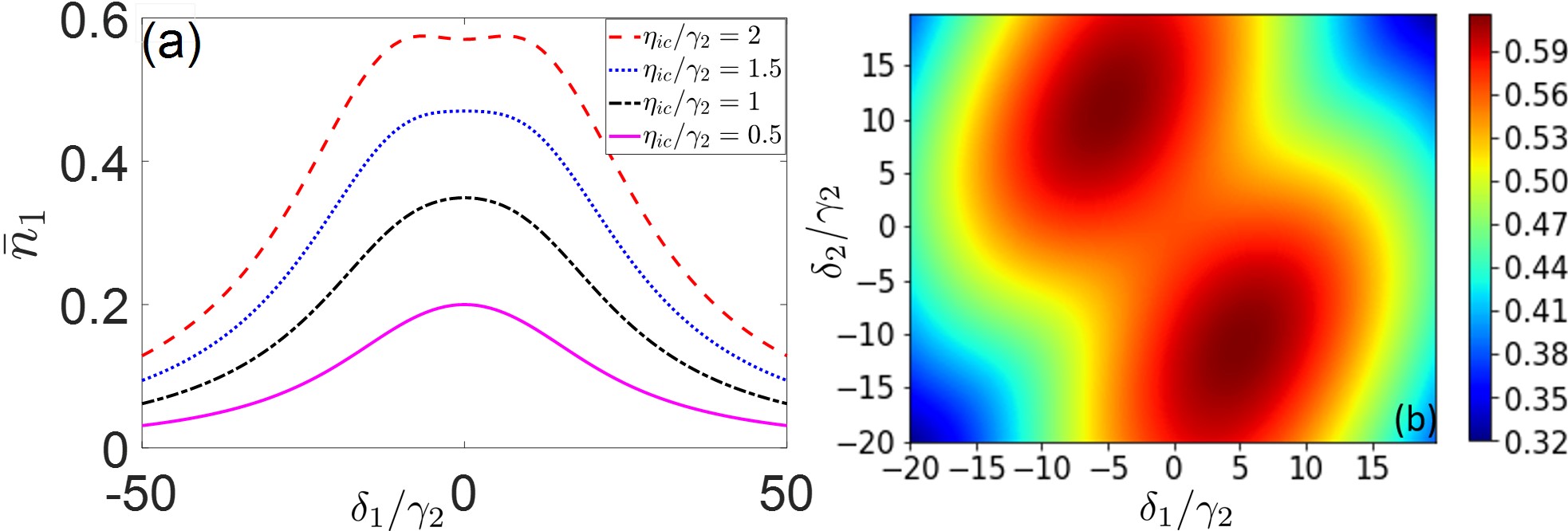}
\end{tabular}
\caption{(a) $ \bar n_1$ is plotted as a function of $\delta_1$ (in unit of $\gamma_2$) for $\eta_{ic} = 2 \gamma_2$ (dashed), $\eta_{ic}=1.5 \gamma_2$ (dotted), $\eta_{ic} = 1 \gamma_2$ (dashed-dotted), $\eta_{ic}=0.5 \gamma_2$ (solid), for $\delta_2 = 0$, $g_1=g_2= 10 \gamma_2$, $\gamma_1 = \gamma_2$ and $\kappa_1 = \kappa_2 = \gamma_2$.
(b) The surface plot exhibits the variation of $ \bar n_1$  as a function of both $\delta_1/\gamma_2$ and $\delta_2/\gamma_2$ for  $\eta_{ic}=2 \gamma_2$.  The surface plot of $\bar n_2$ for  $\eta_{ic}=2 \gamma_2$ will be same as in (b);  but with  $\delta_1$- and $\delta_2$-axis being exchanged.}
 \label{fig 6}
\end{figure}

\begin{figure}
\centering
\vspace{.25in}
  \begin{tabular}{@{}cc@{}}
\includegraphics[width=0.9\linewidth]{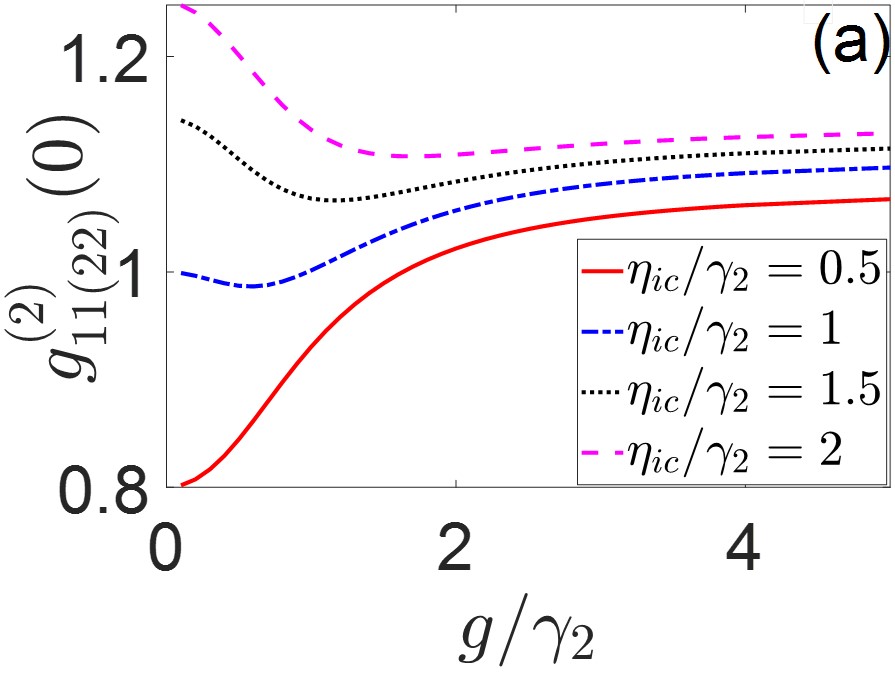}\\
\includegraphics[width=0.9\linewidth]{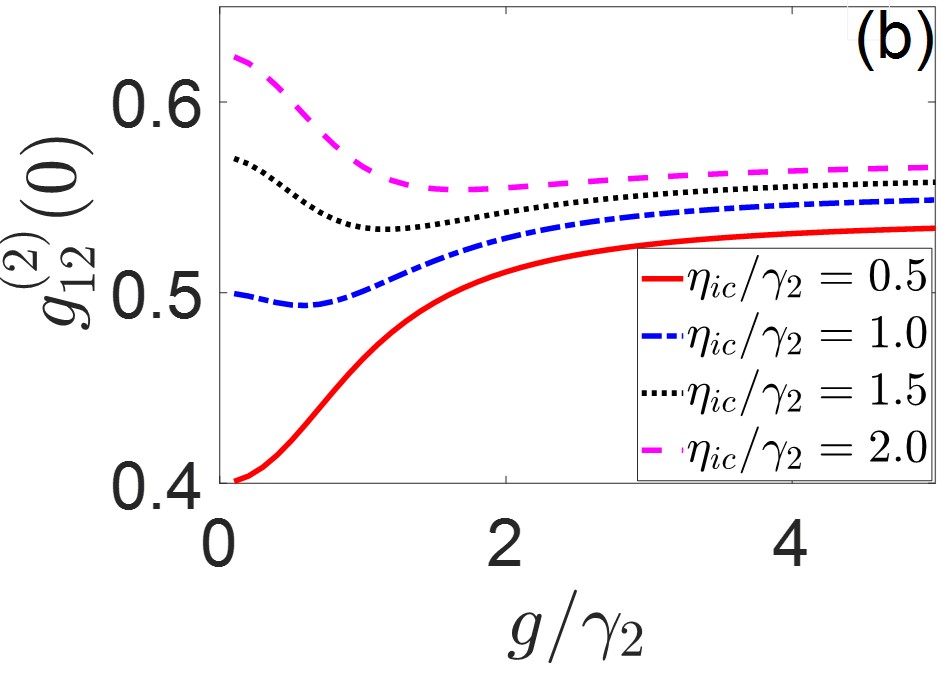}
\end{tabular}
\caption{(a) Equal-time intra-mode HBT correlation functions $g^{(2)}_{11(22)}(0)$  and (b) inter-mode $g^{(2)}_{12}(0)$ are plotted as a function of $g/\gamma_2$ for  different incoherent pump strengths $\eta_{ic}= 0.5 \gamma_2$ (solid), $\eta_{ic}=1 \gamma_2$ (dashed-dotted), $\eta_{ic}=1.5 \gamma_2$ (dotted) and $\eta_{ic}=2 \gamma_2$ (dashed). All other parameters are same as in Fig.~\ref{fig 4} .}
 \label{fig 5}
\end{figure}

\begin{figure}
\centering
\vspace{.25in}
  \begin{tabular}{@{}cc@{}}
\includegraphics[width=0.9\linewidth]{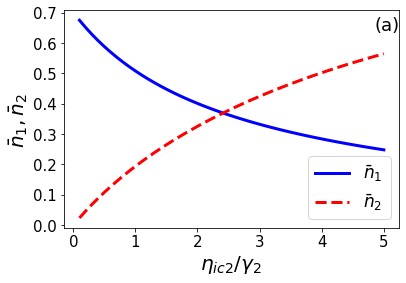}\\
\includegraphics[width=0.9\linewidth]{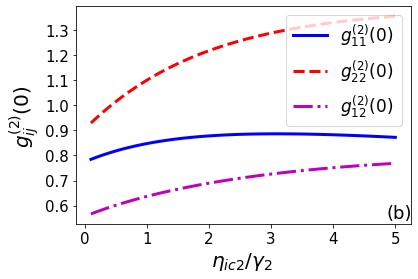}\\
\end{tabular}
\caption{(a) $ \bar n_1$ (solid) and $\bar n_2$
(dashed), (b) $g^{(2)}_{11}(0)$ (solid) , $g^{(2)}_{22}(0)$ (dashed) and $g^{(2)}_{12}(0)$ (dashed-dotted) are plotted as a function of $\eta_{ic2}/\gamma_2$ for $\delta_1=\delta_2=0, \eta_{ic1}/\gamma_2=2, \gamma_1/\gamma_2= 0.1, g_1/\gamma_2 = 1.0, g_2/\gamma_2=3$  }
 \label{figshelving}
\end{figure}

\begin{figure}
\centering
\vspace{.25in}
  \begin{tabular}{@{}cc@{}}
    \includegraphics[width=1\linewidth]{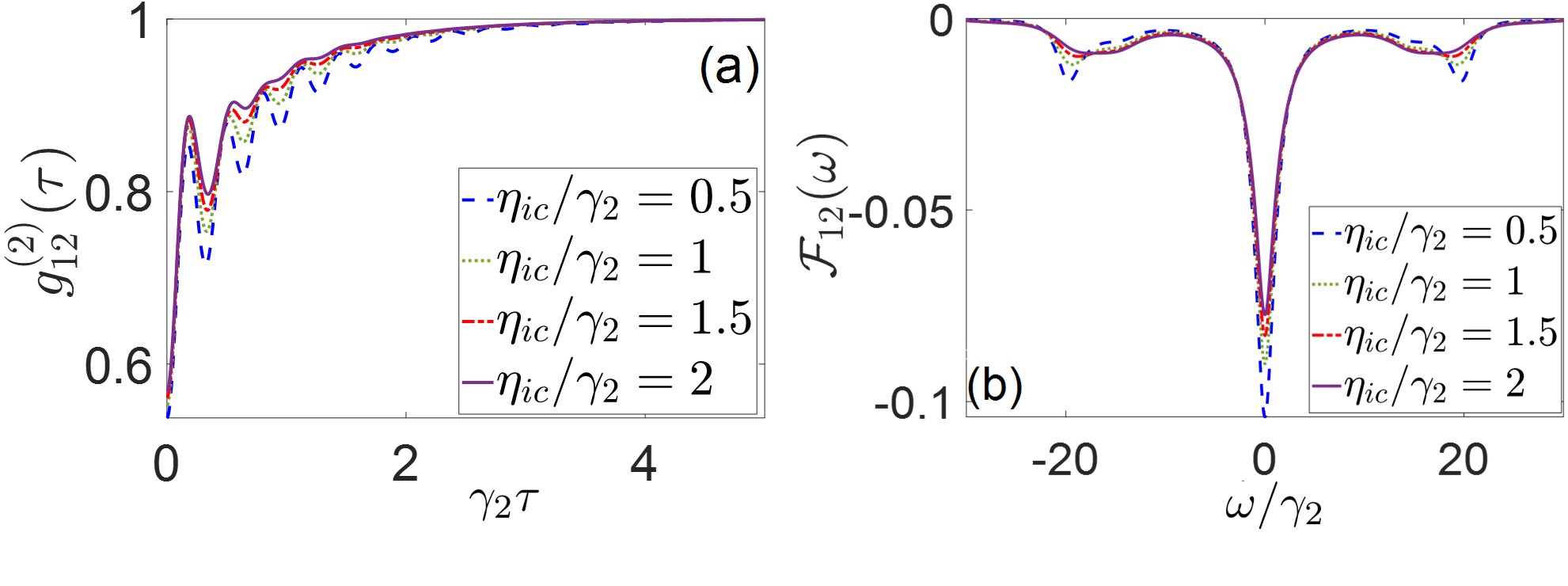}
     \end{tabular}
\caption{(a) The inter-mode HBT correlation function $g^{(2)}_{12}(\tau)$ is plotted against time-delay $\tau$ (in unit of $\gamma_2$) and (b) the spectrum $\mathcal{F}_{12}(\omega)$ of the correlation is plotted against frequency $\omega/\gamma_2$ for $\gamma_1 = \gamma_2$, $\eta_{ic}=0.5 \gamma_2$ (dashed), $\eta_{ic} =  \gamma_2$ (dotted), $\eta_{ic}=1.5 \gamma_2$ (dashed-dotted), $\eta_{ic}=2 \gamma_2$ (solid) keeping the light-matter coupling $g_1=g_2=10 \gamma_2$ fixed. The rest of the parameters are same as in Fig. \ref{fig 4}. }
 \label{fig 7}
\end{figure}
 
  \begin{figure}
\centering
\vspace{.25in}
  \begin{tabular}{@{}cc@{}}
\includegraphics[width=1\linewidth]{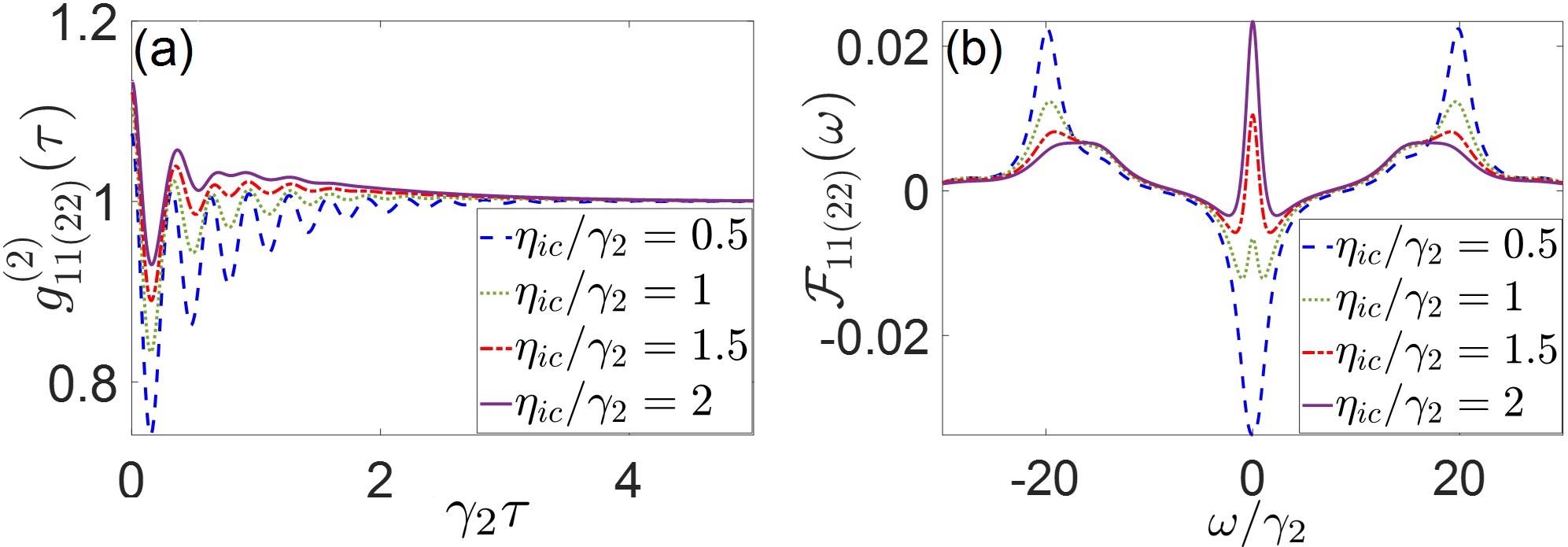}
     \end{tabular}
 \caption{(a) The intra-mode  HBT functions $g^{(2)}_{11}(\tau)=g^{(2)}_{22}(\tau)$ are plotted against time delay $\tau$ (in unit of $\gamma_2^{-1}$) and (b) the spectrum ${\mathcal F}_{ii}(\omega)$ is plotted against frequency $\omega$ (in unit of $\gamma_2$), for $\gamma_1 = \gamma_2$,  $\eta_{ic}=0.5 \gamma_2$ (dashed), $\eta_{ic}= \gamma_2$ (dotted), $\eta_{ic}=1.5 \gamma_2$ (dashed-dotted), $\eta_{ic}=2 \gamma_2$ (solid) with $g_1= g_2 = 10 \gamma_2.$ The rest of the parameters are same as in Fig. \ref{fig 4}. }
  \label{fig 8}
\end{figure}

\begin{figure*}
\centering
\vspace{.25in}
  \begin{tabular}{@{}cc@{}}
\includegraphics[width=0.7\linewidth]{ 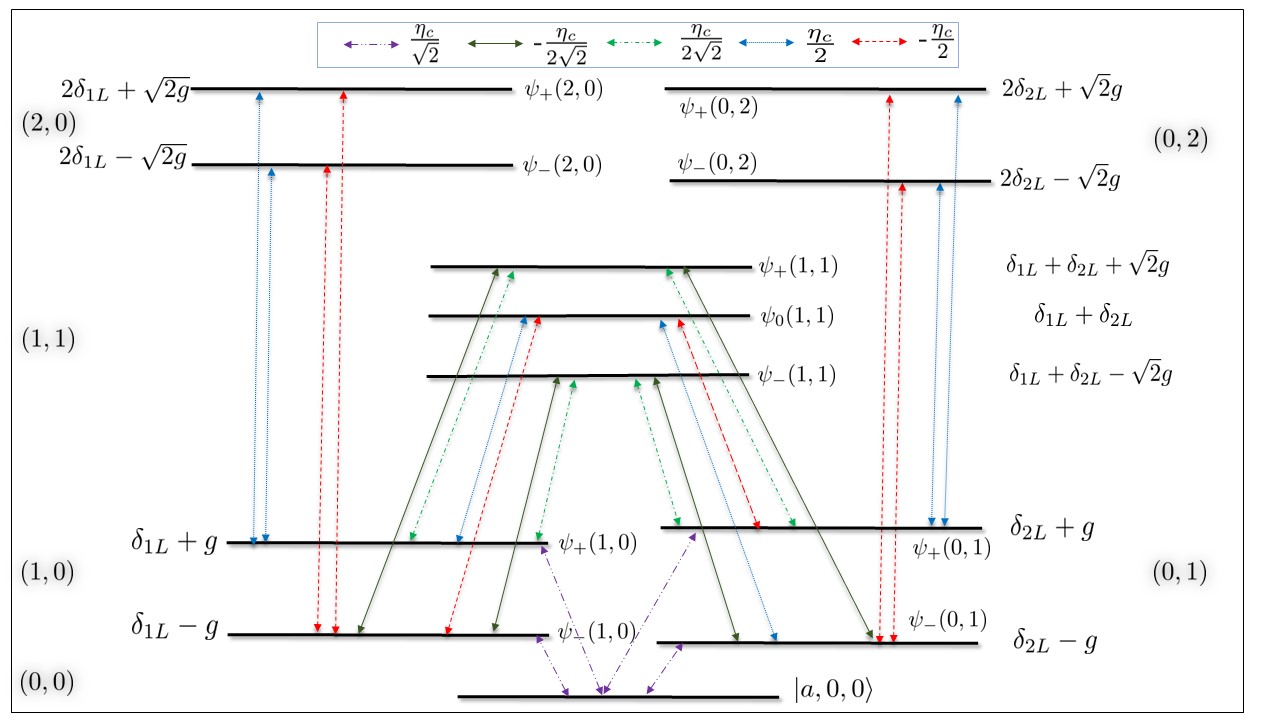}\hfill
     \end{tabular}
 \caption{Dressed-state level diagram showing couplings due to the  coherent drives only for $\eta_{c1} = \eta_{c2} = \eta_{c}$ . The  drives connect the  dressed states from different photon sectors and the transition amplitude for each of the transitions is mentioned.} 
  \label{fig 9}
\end{figure*}


\begin{figure}
\centering
\vspace{.25in}
  \begin{tabular}{@{}cc@{}}
\includegraphics[width=0.9\linewidth]{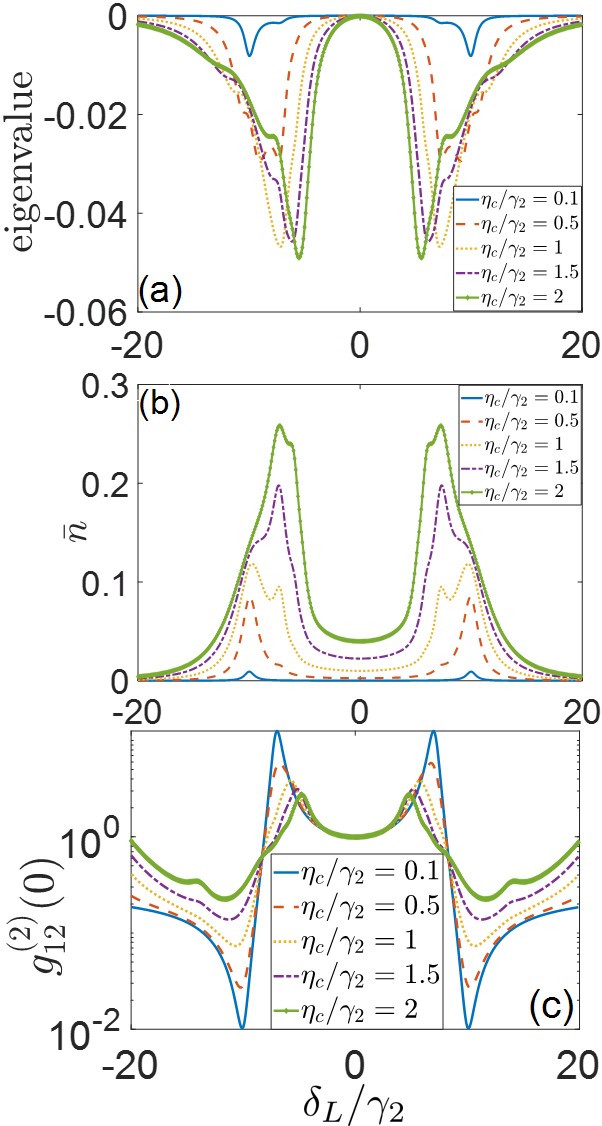}
     \end{tabular}
 \caption{The smallest eigenvalue of $\rho_{f}^{\rm PT,2}$ (a), $\bar n_2=\bar n_1=\bar n$ (b) and $g^{(2)}_{12}(0)$ (c) are plotted against $\delta_{1L}=\delta_{2L}=\delta_{L}$ (in unit of $\gamma_2$) for  $\eta_c=0.1 \gamma_2$ (solid),    $\eta_c=0.5 \gamma_2$ (dashed), $\eta_c=1 \gamma_2$ (dotted),  $\eta_c=1.5 \gamma_2$ (dashed-dotted)  and $\eta_c=2 \gamma_2$ (diamond marker).    The rest of the parameters are $g=10\gamma_2,\gamma_1=\gamma_2,\kappa=\gamma_2$ with $\eta_{ic}=0$.  } 
  \label{fig 11}
\end{figure}

\begin{figure}
\centering
\vspace{.25in}
  \begin{tabular}{@{}cc@{}}
\includegraphics[width=0.9\linewidth]{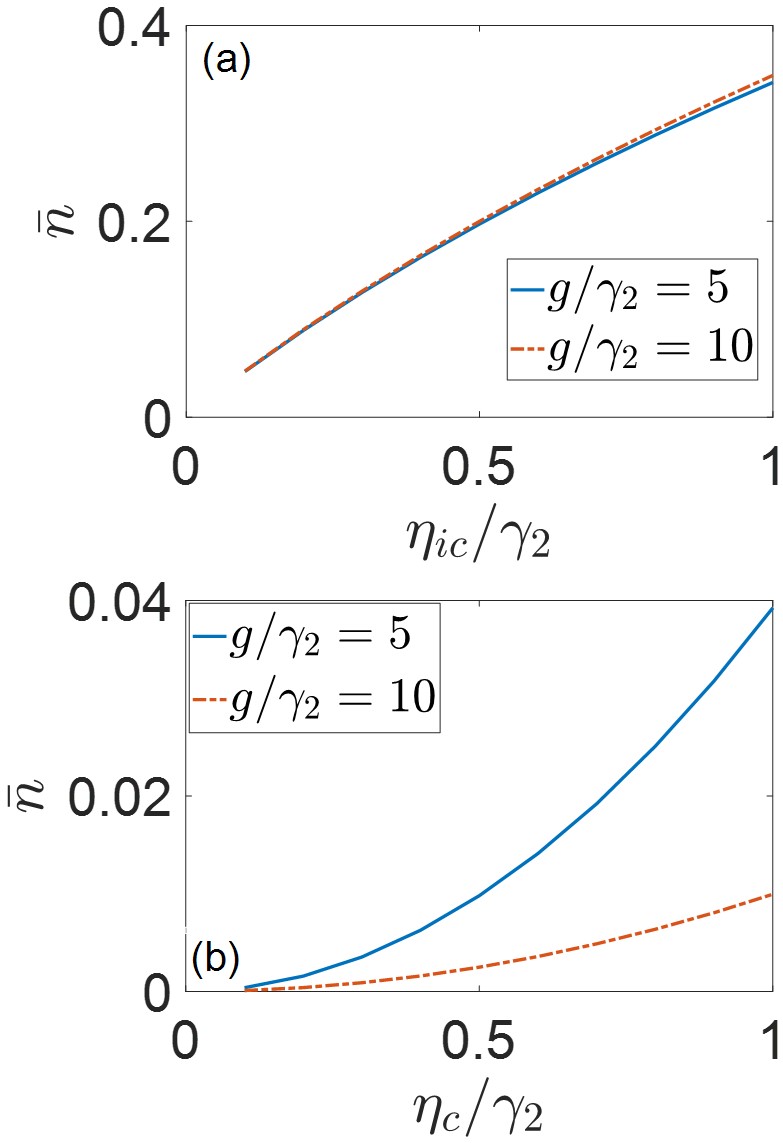}\\
     \end{tabular} 
     \caption{ $\bar{n}$  vs. $\eta_{ic}/\gamma_2$ with $\delta_{1}=\delta_{2}=0$ (a) for $\eta_c=0$; and $\bar{n}$  vs. $\eta_{c}/\gamma_2$ with $\delta_{1L}=\delta_{2L}=0$ (b) for $\eta_{ic}=0$ with $g_1=g_2=g=5\gamma_2$ (solid) and $g=10 \gamma_2$ (dashed-dotted). The other common parameters for both (a) and (b) are , $\gamma_1 = \gamma_2$ and $\kappa_1 = \kappa_2 = \gamma_2$. }
  \label{fig 13}
\end{figure}

\begin{figure*}
\centering
\vspace{.25in}
  \begin{tabular}{@{}cc@{}}
\includegraphics[width=0.31\linewidth]{ 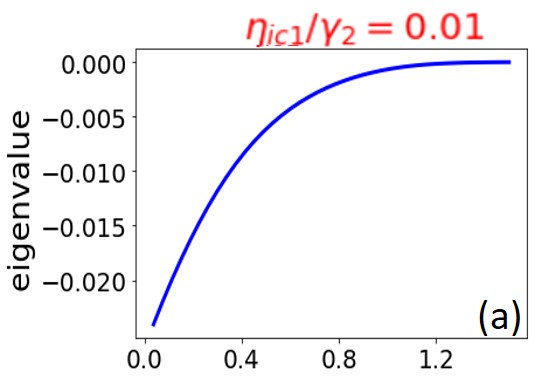}\hfill
\includegraphics[width=0.3\linewidth]{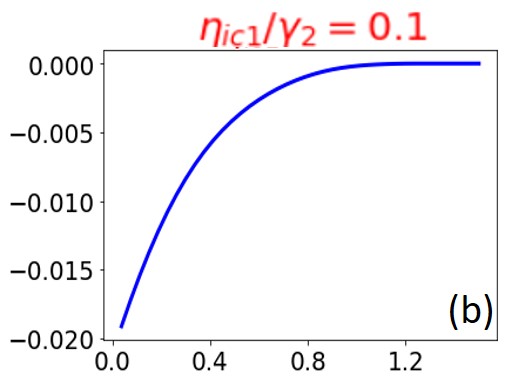}\hfill
\includegraphics[width=0.3\linewidth]{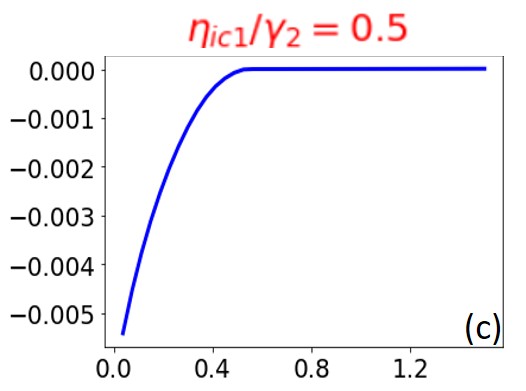}\\
\includegraphics[width=0.3\linewidth]{ 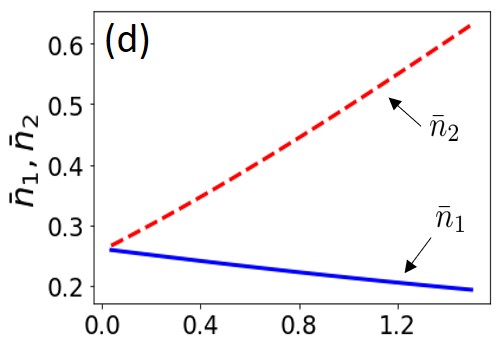}\hfill
\includegraphics[width=0.278\linewidth]{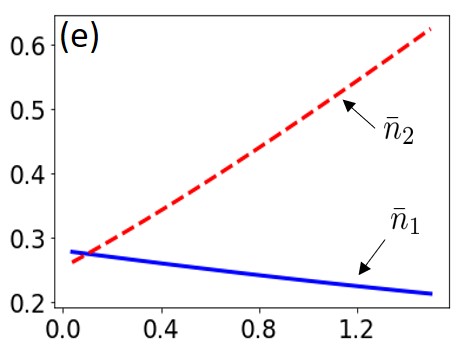}\hfill
\includegraphics[width=0.28\linewidth]{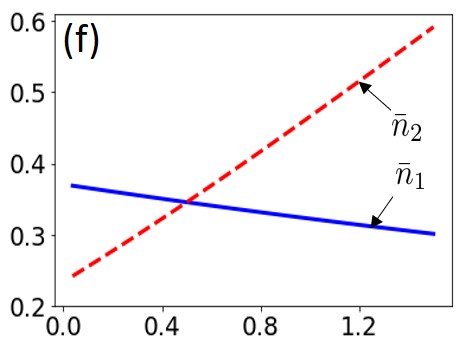}\\
\includegraphics[width=0.3\linewidth]{ 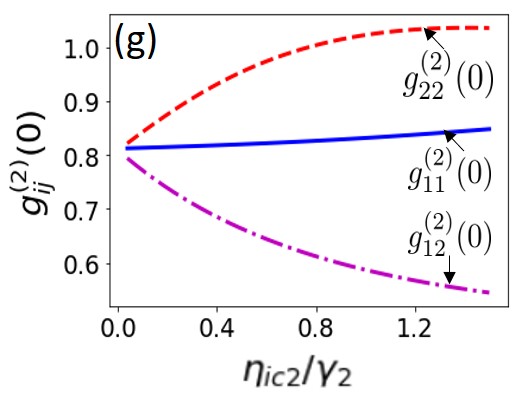}\hfill
\includegraphics[width=0.27\linewidth]{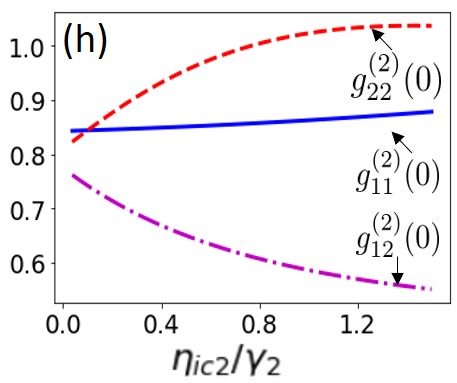}\hfill
\includegraphics[width=0.27\linewidth]{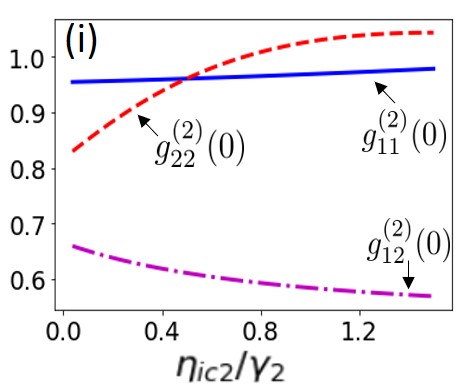}\\
     \end{tabular}
 \caption{ The smallest eigenvalue of $\rho_{f}^{\rm PT,2}$ (a, b c), $\bar n_1$, $ \bar n_2$ (d, e, f) and $g^{(2)}_{ij}(0)$ (g,h i) are plotted against  $\eta_{ic2}/\gamma_2$ for  $\eta_{ic1}=0.01\gamma_2$ (a, d, g),    $\eta_{ic1}=0.1\gamma_2$ (b,e,h),  $\eta_{ic1}=0.5\gamma_2$ (c,f,i), keeping $\delta_{1L}=\delta_{2L}= g/\sqrt{2}$ (in the unit of $\gamma_2$).  The rest of the parameters are same as in Fig. \ref{fig 11}.}
  \label{fig en}
\end{figure*}

First we consider the case when the three-level atom inside the cavity is  pumped incoherently with no coherent drive being present.  The level diagram with only incoherent couplings between the dressed states is schematically shown in Fig. \ref{fig 2} (ii) and (iii). Figure ~\ref{fig 3} shows the convergence of the values of the average photon numbers $\bar n_1=\bar n_2=\bar n$, against the variation of the Fock basis size $N_{max}$ for a chosen set of equal parameters for the two cavity modes. We find that the results converge for $N_{max} \ge 5$ for the chosen parameter set as mentioned in the figure caption. For the strong-coupling regime limited by $g_1 \le 10$ and $g_2 \le 10$;  and $\eta_{ic1} \le 2$ and $\eta_{ic2} \le 2$, we can safely restrict Fock states per cavity mode at $N_{max} = 6$.  

Figure~\ref{fig 4} demonstrates that the dual incoherent pumping leads to the thresholdless two-mode lasing action. From Fig.~\ref{fig 4}(a), one can notice that as the atom-cavity coupling strengths $ g_1= g_2= g$ increase for equal and fixed incoherent pump strengths $\eta_{ic1}=\eta_{ic2}=\eta_{ic}$  the average photon numbers $\bar n_1=\bar n_2 = \bar n$ first start growing and then saturate. Larger the value of $\eta_{ic}$, the larger is the saturation limit revealing that the dual incoherent pumping leads to amplification in both the modes. The surface plot in Fig.~\ref{fig 4}(b) shows the variation of $\bar n_1$  against both $g_1$ and $g_2$. The surface plot of $\bar n_2$ as a function of both  $g_1$ and $g_2$ is same as in   Fig.~\ref{fig 4}(b), but the two axes should be interchanged.   These plots suggest that one can control the relative photon numbers in the two modes by changing the two atom-cavity coupling parameters. Below the saturation limit of $g_i$ ($i=1,2$), the average photon number in the $i$th mode is directly proportional to $g_i$. If we switch off one of the incoherent pumps, then the system reduces to an effective two-level atom pumped incoherently inside a single-mode cavity, leading to single-mode CQED laser. 

In Fig.~\ref{fig 6}(a), we plot $\bar{n}_1$ vs. $\delta_1$ for $\delta_2 = 0$ for four different values of $\eta_{ic}$ as mentioned in the legend of the figure. It is important to note that, as we increase $\eta_{ic}$ above a certain value ($\eta_{ic}\ge 1.5$),   a single-peak structure is split into  a double peak one.   Figure~\ref{fig 6}(b) displays the variation of $\bar n_1$ as a function of both $\delta_1$ and $\delta_2$  for $\eta_{ic}= 2$, clearly depicting the two-peak structure. The variation of $\bar n_2$ against $\delta_1$ and $\delta_2$ for $\eta_{ic}= 2$ is same as in Figure~\ref{fig 6}(b), but the  $\delta_1$- and $\delta_2$-axis should be interchanged. From this surface   plots, the positions of the two peaks for $\bar{n}_1$ are detected to be $(g/2, -\sqrt{2} g)$ and $(-g/2, g)$ while those for $\bar{n}_2$ are $(\sqrt{2} g, -g/2)$ and $(-g, g/2)$, indicating that probably the lasing in the two modes occur due to stimulated transitions between the dressed states. For weak incoherent pumping, the amplification or lasing occurs primarily due to the transitions in the lower photon sectors, namely, (0,0)-, (1,0)- and (0,1)-sectors. In these sectors, the cavity-field stimulated transitions $\mid b, 0, 0 \rangle \rightarrow \mid a, 1, 0 \rangle $ and $\mid c, 0, 0 \rangle \rightarrow \mid a, 0, 1 \rangle $   to the two modes occur independently in two non-coupled two-level systems. The intensity of the emitted radiation is broadened due to strong CQED coupling giving the linewidth $\sim 2 g + \gamma + \kappa$ which is 22 for the parameters chosen. When incoherent pumping strength is increased beyond 1.5, the higher photon sectors (0,2),(2,0) becomes involved in the stimulated transition process. As shown in the bare- and dressed-state level diagram of Fig.\ref{fig 2}, all three bare- or dressed-states become involved in the process for higher photon-sectors, leading to two-peak structure in the radiation intensity.

In Fig.~\ref{fig 5}(a) equal-time Hanbury Brown-Twiss (HBT) intra-mode correlation function $g_{11(22)}^{(2)}(0)$ is plotted against  coupling parameter ${g_1 = g_2 = g}$, for four different values of $\eta_{ic}$:  0.5 (solid), 1.0 (dashed), 1.5 (dotted), 2 (dashed-dotted). If all other parameters corresponding to both the modes are equal, then $g_{11}^{(2)}(0) = g_{22}^{(2)}(0)$. Figure~\ref{fig 5}(b) shows equal- time  inter-mode HBT correlation function $g_{12}^{(2)}(0)$ for the same parameters  . For small values of $\eta_{ic}$ and atom-cavity coupling constants being not very high, both inter and intra-mode HBT correlations show nonclassical behavior characterized by $g_{11(22)}^{(2)}(0)<1, g_{12}^{(2)}(0)<1$. As $\eta_{ic}$ increases for fixed values of $g$,  the inter-mode correlation remains  nonclassical ($g_{12}^{(2)}(0)<1$), while the intra-mode correlation approaches thermal limit, i.e., $1 < g_{11(22)}^{(2)}(0) < 2$. It is interesting to note that,  the cavity photons of the same mode are most likely to  appear in pairs, particularly for strong incoherent pumping,  In contrast, if a photon in one mode is created in the case of weak incoherent pumping, a photon in the other mode can not be created at the same time. This feature may be explained from the point of view of an interplay between the coherent cavity-field Rabi dynamics and the incoherent processes as schematically depicted in the level diagram of Fig.\ref{fig 2}. When the atom is initially in the ground state $\mid a, 0, 0 \rangle$ with the two cavity fields in the vacuum, the incoherent pumping can raise the atom  to either of the excited states $\mid b, 0, 0 \rangle$ or $\mid c, 0, 0 \rangle$. Suppose, the atom is excited to $\mid b, 0, 0 \rangle$, Then the atom can either emit a photon in cavity mode 1 or into the vacuum by atomic decay. If it emits a photon into the cavity mode 1, then it will find itself in state $\mid a, 1, 0 \rangle$ until the photon leaks out of the cavity with the characteristic time scale $\kappa^{-1}$. In both the cases of atomic decay and cavity decay, the system will be reset to its initial state  $\mid a, 0, 0 \rangle$. Now, if the incoherent pumping strength is weak, then it is most likely that the system will be reset to its initial state  rather than being promoted to the next higher photon sector. In the strong-coupling CQED regime with one photon in mode 1 and no photon in mode 2, there will Rabi oscillations between the two atomic states $\mid a \rangle$ and $\mid b \rangle$ before the atom or the photon decays.   Thus, for weak incoherent pumping, until and unless the atom is excited to $\mid c, 0, 0 \rangle$ from the reset state  $\mid a, 0, 0 \rangle$ by incoherent processes, a photon in the mode 2 will not appear. This is the physical origin of the inter-mode photon-photon anti-bunching. 

 We next show how the results are affected when the two spontaneous emission  linewidths are widely different, that is, when  the lifetime of one of the excited states is much larger than that of the other. This question arises because in experiments with ultracold two-electron atoms such as Yb, Sr, etc., one can populate metastable $P$-excited states (whose life times can be of the order a second) by optical pumping. So, the question naturally arises how the two-mode lasing action and quantum correlations as discussed above will change if  a $V$-type three-level atom with one metastable excited state is used in our proposed bimodal single-atom CQED laser.  Such $V$-type three-level atom had been used earlier for studying shelving effect and quantum jumps \cite{cohen1998atom,dehmelt1975proposed, PhysRevLett.54.1023} in spectroscopy. In Fig. \ref{figshelving}, we show the results considering the case $\gamma_1 = 0.1 \gamma_2$, that is, the lifetime of the state $\mid b \rangle$ is 10 times larger than that of $\mid c \rangle$. Then, considering that the two cavity-mode volumes are almost same, one should expect $g_1/g_2 \simeq \sqrt{\gamma_1/\gamma_2}$ since atomic damping constant is proportional to the square of the matrix element of transition dipole moment while the Rabi frequency is proportional to  the dipole moment matrix element. We therefore set $g_2/g_1 = 3$ in Fig.  \ref{figshelving} which shows how the variation of the relative strengths of the dual incoherent pumps can  influence the relative intensity and the photon statistics of the two cavity fields.  It is interesting to note that,  as Fig. \ref{figshelving} (a) reveals, with the increase of the incoherent pumping rate $\eta_{ic2}$ keeping $\eta_{ic1}$ fixed, the field  intensity (average photon number) of mode 2 increases at the cost of the field intensity of mode 1, suggesting that the tuning of the dual incoherent pumping can lead to photon population transfer between the two cavity modes.

Fig.~\ref{fig 7}(a) shows the variation of $g_{12}^{(2)}(\tau)$ against $\tau$ while Fig.~\ref{fig 7}(b) exhibits the spectrum $\mathcal{F}_{12}(\omega)$  as a function of frequency $\omega$ for four different values of $\eta_{ic}$ as mentioned in the legend.  As $\tau\rightarrow \infty$, $g_{12}^{(2)}(\tau)$ reaches the value of unity implying that the two modes become uncorrelated in the long time delay limit. For low values of $\eta_{ic}$, $g_{12}^{(2)}(\tau)$ exhibits oscillatory decay from nonclassical domain  $(g_{12}^{(2)}(\tau) < 1)$ towards coherent limit $(g_{12}^{(2)}(\tau) \rightarrow 1)$. As we increase $\eta_{ic}$, the amplitude of the oscillations gradually diminish. These oscillations are reminiscent of quantum beats and indicative of atomic coherence. The Fourier transform of $g_{12}^{(2)}(\tau) - 1$, that is,  ${ \cal F}_{12}(\omega)$ as a function of $\omega$ as shown in Fig.\ref{fig 7}(b) reveals that there is one dominant frequency of oscillations which is $2g$ in this case. We recall the the vacuum-field Rabi splitting of two-level Jaynes-Cummings model is $2g$. This reveals that for low incoherent pumping, the photons are generated due to stimulated transitions in the (1,0)- and (0,1)-photon sectors. Remarkably, the anti-correlation between the two field modes $g_{12}^{(2)}(\tau)- 1 < 0$ is also reflected in the ${\cal F}_{12}(\omega)$ vs. $\omega$ plots. This can be ascertained from the observation that  the central peak of ${\cal F}_{12}(\omega)$ at $\omega = 0$  is negative. We have checked that as we gradually  increase the strength of the incoherent pump, a new smaller dip appears at $\omega = \sqrt{2}g$ in the  ${\cal F}_{12}(\omega)$ vs. $\omega$ plot. The dressed-state energies of (1,1)-photon sector for $\delta_1 = \delta_2 =0$ are $-\sqrt{2} g$, 0 and $\sqrt{2}g$ and for (2,0) and (0,2) photon sectors it is  $\pm \sqrt{2}g$. But (1,1) photon sector will not take part in this process as discussed. This indicates that for relatively higher incoherent pump strength, the oscillations in the photon-photon correlations correspond to Rabi oscillations in these two-states CQED dynamics. 
 
 We display the variation of $g_{11(22)}^{(2)}(\tau)$ as a function of $\tau $ against in Fig.~\ref{fig 8} (a) for four different values of $\eta_{ic}$. The corresponding spectrum is shown in Fig.~\ref{fig 8} (b). As in the case of inter-mode correlations, the intra-mode correlation too exhibits oscillatory decay. However, in contrast to inter-mode case, the intra-mode correlations become predominantly classical  $(g_{12}^{(2)}(\tau) \ge 1 )$ for  $\eta_{ic} > 1$ while they are mostly nonclassical for  $\eta_{ic} \le 1$. These correlation properties are also manifested in the spectral profiles in that the central peak of ${\cal F}_{ii}(\omega) $ at $\omega = 0$ is positive  or negative for the classical or nonclassical correlations, respectively.

\subsection{The effects of  coherent drives: Photon-Photon entanglement}

Here we calculate the lowest eigenvalue of partially-transposed matrix of the reduced two-mode field density matrix as discussed in the subsection \ref{D} of the previous section to verify the existence of photon-photon entanglement.  The negativity of the eigenvalue is a clear signature of the entanglement.  For our chosen parameter regime and truncated Fock space, the reduced two-mode field density matrix  is of 36$\times$36 dimension. 

Under the two-photon resonance condition $\delta_1 = \delta_2= \delta$, in a frame rotating with the driving laser frequencies, 
the eigenvalues of the three dressed states are given by  
\begin{align*}
E_0/{\hbar} &= n_1 \delta_{1L}+ n_2\delta_{2L} - \delta~, \\
E_{\pm}/{\hbar} &= n_1 \delta_{1L} + n_2 \delta_{2L} - \frac{\delta}{2} \pm  \frac{1}{2} \sqrt{ \delta^2 + 4 \left ( n_1 g_1^2 + n_2 g_2^2 \right ) }~,
\end{align*}
where $\delta_{1L} = \omega_1 - \omega_{1L} = \delta_1 + \Delta_{b}$ and $\delta_{2L} = \omega_2 - \omega_{2L} = \delta_2 + \Delta_c$ are two cavity-drive detuning parameters. The dressed-state level diagram depicting the transitions due to the coherent drives that connect different photon sectors is shown in Fig.~\ref{fig 9} for $\eta_{c1} = \eta_{c2} = \eta_{c}$. 
The  amplitudes of the transitions are mentioned in the diagram. These transition amplitudes are given by
\begin{eqnarray}
\langle \psi_{\alpha}(n_1, n_2) \mid \hbar \eta_c \left (\sigma_1 + \sigma_2 + {\rm H.c.} \right ) \mid \psi_{\beta}(n_1', n_2') \rangle~.
\end{eqnarray}
 where $\alpha, \beta = 0, \pm$, $\mid \psi_{\alpha}(n_1, n_2) \rangle$ and $\mid \psi_{\beta}(n_1', n_2') \rangle$ refer to the dressed states as given 
by Eqs. \ref{eqn psipm} and \ref{eqn psibc}. In the absence of incoherent pumping, as $\eta_c$ increases from zero the coherent drives will couple from lower to higher dressed  states successively as a consequence of the competition between the driving and the damping processes.   The  drives make superposition of the cavity-dressed states. The superposition state is an atom-field entangled state. Since the field is a two-mode one, if the atom is measured in a particular state, the atom-field entangled state will be projected into a two-mode entangled field state.

 We first study the system driven by the dual coherent drives in the absence of any incoherent pump.  The smallest eigenvalue of $\rho_{f}^{{\rm PT,2}}$ as a function $\delta_{2L}=\delta_{1L}=\delta_{L}$ is plotted in
  Fig.~\ref{fig 11} (a). This figure shows that the eigenvalue becomes more negative with  the increase of the coherent drive strength, indicating that the two cavity fields are entangled.  In $\bar{n}_2,\bar{n}_1$  vs. $\delta_{L}$ plots of Fig.~\ref{fig 11}(b), the peak at $\delta_{L}= g = 10$ corresponds to the emissions from (0,1),(1,0) photon sector. An additional peak appears at $\delta_{L} = g/\sqrt{2} \simeq 7.1 $ with the increase  of $\eta_c$ as a signature of the emission from (0,2),(2,0)-photon sector. An emergence of another peak is observed at $\delta_{L} = g/\sqrt{3} \simeq 6 $ as an evidence of probing (2,1) and (1,2) photon sectors. Since both $\bar n_1$ and $\bar n_2$ are equal and appreciably enhanced, it is meaningful to analyse the statistical properties in this parameter regime. In Fig. \ref{fig 11}  most of the dips in the negativity of the eigenvalue correspond to the peaks in the photon number populations in both the modes as can be inferred by comparing  Figs. \ref{fig 11} (a) with \ref{fig 11}(b). In Fig.  \ref{fig 11}(c), $g_{12}^{(2)}(0)$ is plotted as a function of $\delta_L$.  It is to be noted that the inter-mode antibuncing does not always necessarily mean inter-mode  entanglement.
 
 We next compare the effects of dual coherent drives vis-a-vis of dual incoherent pumping on the properties of the cavity fields. Figure \ref{fig 13} illustrates that while only coherent drives with no incoherent pumping can amplify the cavity fields up to certain limit set by the saturation of the atomic populations, the dual incoherent pumping with no coherent drive with all other input parameters remaining same leads to substantially large amplification. Particularly in the regime far below the saturation limit, the amplification due to incoherent pumping is one order of magnitude larger than that due coherent drives only. Figure \ref{fig en} displays comparative results on the effects of coherent vs. incoherent pumping on  the entanglement and photon statistical properties of the two cavity fields. It is important to note the essential role of incoherent pumping in a single-atom  CQED laser. From Fig. \ref{fig en} (g), (h), (i) it is evident that as we increase $\eta_{ic1}$ and $\eta_{ic2}$ to certain extent but much lower than the atomic as well as cavity damping  constants, the nonclassicality of the inter-mode HBT correlations is enhanced. Looking at  Fig. \ref{fig en} (a),  (b), (c), (g), (h) and (i),  one can infer that as the dual incoherent pumping increases up to about $0.4 \gamma_2$, both  two-mode entanglement and the nonclassicality persist.  In all these parameter regimes, as Figs. \ref{fig en} (d),  (e) and (f) show,  a moderate amount of incoherent pumping ensure an  appreciable amount of photon population in the two modes so that the detection of photons from each mode becomes highly probable. It is indeed important to note that incoherent pumping is not only essential for amplifying the photon populations but also important to enhance the nonclassicality, though  the entanglement is produced only when  coherent drives are used. In this context, it may be worthwhile to mention that the  competition between incoherent pumping and coherent cavity QED has been theoretically shown to lead to synchronization between two atomic ensembles inside a cavity \citep{PhysRevLett.113.154101}. 
 
Before ending this section, it is worthwhile to make a few comments about the origin of the two-mode entanglement discussed above.  As mentioned before, the origin  can be traced to the dressing of the cavity-dressed states by  the coherent drives. The possibility of generating atom-photon entangled states, Schroedinger cat states of cavity fields and two-mode entangled field states by  driving two-level atoms with a strong coherent field inside a single-mode cavity had been discussed by Solano, Agarwal and Walther \cite{PhysRevLett.90.027903} about two decades ago. They considered the dressed states of atoms due to the strong classical drive and then these atomic dressed states are then superposed  by the  cavity field, producing atom-cavity dressed states and thus leading to the entanglement.  In our case, the two strong cavity fields first dress the three-level atom, leading to  bimodal atom-cavity dressed states which are then further dressed by the coherent drives. Thus tripartite entangled states involving the atom and the two cavity modes are produced. Now, if the $V$-type three-level atom is measured in the ground state, the tripartite entangled state will be projected into a two-mode entangled field state.  In fact, generation of two-mode entangled fields in the microwave frequency domain using a bimodal superconducting cavity and circular Rydberg atoms  had been experimentally demonstrated by Heroche's group in 2001 \cite{haroche:pra64:1050:2001}. In that experiment, the cavity was driven by  excited Rydberg atoms that had passed through the cavity one by one. In 2005, Xiong, Scully and Zubairy \cite{xion:prl94:023601:2005} had theoretically shown that a "Correlated-spontaneous emission laser" using an ensemble of cascade or $\Lambda$-type three-level atoms driven by two classical coherent fields inside a doubly resonant cavity can produce two-mode entangled radiation with large number of photons in each mode.

\section{Conclusions}\label{sec4}
In conclusion, we have proposed a single-atom CQED laser that can generate nonclassically correlated and entangled photon pairs. We have shown that when the two cavity modes are incoherently pumped without any coherent drive, the laser produces amplified signals for both the modes with inter-mode nonclassical photon-photon correlation  but with no photon-photon entanglement. The nonclassical behavior is manifested though inter-mode photon-photon anti-bunching which is shown to result from an interplay between the incoherent processes and cavity-field induced coherent processes. When coherent atomic drives are used with small or no incoherent pumping, the laser generates inter-mode entangled photon pairs. We have demonstrated that the entanglement arises due to the dressing of the cavity-dressed states by the coherent drives.  As elaborated in our paper, not only the intensities  of the  generated fields,  but also the photon-photon correlations and entanglement can be controlled at a microscopic level by varying the incoherent pumping rates, coherent drive strengths and detuning parameters, thus providing a wide range of functionality of our proposed laser system. We have further shown the nontrivial effects of dual incoherent pumping on the amplification of the small signal of entangled photon pairs which are primarily generated due to coherent drives. Apart from the generating two-mode nonclassical HBT correlations and entanglement, our proposed two-mode CQED laser is expected to deliver two-mode quantum light at a single  or a few photon level with inter-mode squeezing in various quadrature variables. We hope to address this problem of two-mode squeezed light generation by CQED in our future communication.    

Our proposal may be implemented experimentally by combining the currently available technologies of cavity QED and atom traps. For example,  an atom trap containing a single bosonic atom may be loaded inside a bimodal cavity. The three levels of the atom may be chosen in the following way: Let the atom  in the ground electronic $S$-state  be prepared in the hyperfine spin state $\mid F, m_F = 0 \rangle$, where $F$ is the hyperfine spin quantum number and $m_F $ denotes the magnetic hyperfine spin quantum number. The two excited ($P$) states be represented by $\mid F', m_F' = - 1\rangle$ and $\mid F', m_F' = 1\rangle$, where the excited hyperfine spin $F' \ge 1$. Now, if the two cavity modes are chosen to be left and right circularly polarized fields then our proposed one-atom laser will produce polarization-entangled photon pairs with antibunching. This may be realized with a doubly resonant cavity for which the two cavity axes for the two modes may coincide as shown in Fig.\ref{fig 1} (ii).  As regards the implementation of incoherent optical pumping, Kimble's group at Cal-Tech has already experimentally demonstrated such pumping in CQED  using incoherent Raman transitions in a single-atom far-off resonance trap (FORT) loaded inside a cavity \cite{thesis_boozer2005raman}. In the single-mode CQED laser with a single trapped Cs atom reported in Ref. \cite{Nature1}, the laser output was available for a short duration limited  by the lifetime of the trap which was 50 ms. This time limitation may be overcome by using single-ion trap  instead of a single-atom trap in  CQED regime \cite{walther:epl:1997, blatt_PhysRevLett.114.023602}. Because. the lifetime of an ion-trap is quite large. Furthermore, an ion trap may offer some advantages as far as coherent or incoherent optical pumping is concerned as the stability of an ion-trap does not depend on the internal electronic states of the ion unlike that of an atom-trap. Therefore, ion-trap CQED may be an interesting research direction for generating nonclassical  and entangled photon pairs for practical applications.

\bibliography{cite}
\end{document}